\documentstyle[preprint,tighten,aps,epsfig]{revtex}

\def\lapprox{\mathrel{\mathop
  {\hbox{\lower0.5ex\hbox{$\sim$}\kern-0.8em\lower-0.7ex\hbox{$<$}}}}}
\def\gapprox{\mathrel{\mathop
  {\hbox{\lower0.5ex\hbox{$\sim$}\kern-0.8em\lower-0.7ex\hbox{$>$}}}}}

\begin{document}
\author{S. Degl'Innocenti$^{1,2}$, G. Fiorentini$^{3,4}$, B. Ricci$^{3,4}$ and F.L. Villante$^{3,4}$ }

\address{	$^{1}$Dipartimento di Fisica dell'Universit\`a di Pisa,
       via Buonarroti 2, I-56126 Pisa, Italy\\
       $^{2}$Istituto Nazionale di Fisica Nucleare, Sezione di Pisa,
       via Buonarroti 2, I-56126 Pisa, Italy.
       $^{3}$Istituto Nazionale di Fisica Nucleare, Sezione di Ferrara,
        via Paradiso 12, I-44100 Ferrara, Italy, \\
        $^{4}$Dipartimento di Fisica dell'Universit\`a di Ferrara,
        via Paradiso 12, I-44100 Ferrara, Italy
}

\preprint{\vbox{\noindent
          \null\hfill  INFNFE-12-03}}

\title{The $^{14}N(p, \gamma)^{15}O$ 
reaction, solar neutrinos and the age of the
globular clusters}

\maketitle

\begin{abstract}

We discuss implications of a new measurement of $^{14}N(p, \gamma)^{15}O$ 
 concerning solar neutrinos, solar models and globular cluster dating.
Predictions for the gallium and chlorine experiments are reduced by 2 and 0.1 SNU respectively.
Predictions for helioseismic observables are unchanged within uncertainties.
The age of globular clusters as deduced from the Turn-Off luminosity
 is increased by about 0.7 Gyr.

\end{abstract}

\subsection{Introduction}

The last few years  have  presented us  with a remarkable
progress in understanding
hydrogen burning in stars: the solution of the solar neutrino 
puzzle provided by SNO \cite{sno} and KamLAND \cite{kamland} experiments has allowed a precise
determination of the $^8B$ neutrino flux, in  agreement with 
the theoretical prediction of Standard Solar Model (SSM) calculations.
The predicted signals for gallium and chlorine detectors are also 
in good agreement with SSMs, once the survival probability
of electron neutrinos is calculated according to the 
Large Mixing Angle (LMA) solution, see e.g. \cite{roadmap}
\footnote{Actually, there is a slight ($\simeq 2\sigma)$
tension between prediction and measurement for  the  chlorine experiment \cite{roadmap,smirnov}.}.
In this way, the nuclear energy source of the sun,
proposed by Eddington\cite{eddington}, Bethe\cite{bethe} and von Weizs\"{a}cker\cite{von} long ago,
has been checked with observation: the  production
rate of nuclear energy in the sun, as deduced from neutrino observations, agrees
with the observed photon luminosity to within about 20\% \cite{roadmap,beyond03}.

On the laboratory side, recent experiments have provided refined
measurements of  nuclear cross sections relevant
to hydrogen burning: the $^3He(^3He,2p)^4He$ astrophysical
$S$-factor has been measured well in the 
solar Gamow peak with 6\% accuracy \cite{luna33}
and that of  $^7Be(p,\gamma)^8B$ is presently  known with
an accuracy of about 5\% as a result 
of several  recent measurements, see \cite{beyond03}  and refs. therein.

Although rather precise concerning  the $pp$ chain, neutrino observations 
and laboratory experiments shed little light on the role of the CNO bi-cycle in the sun.
This cycle,  producing a tiny ($\approx 1.5\%$ according to SSM) contribution  to the  solar energy output, 
is most important in more advanced burning phases, where it sustains a large fraction of the 
stellar luminosity by hydrogen burning in 
shells as soon as hydrogen is exhausted at the stellar center.

The age of the globular clusters, the oldest systems in the Galaxy, is determined by locating the 
turning point on the Hertzsprung-Russell diagram, i.e. the point which signals hydrogen 
exhaustion at the stellar center.
The efficiency of the CNO cycle is thus expected to be relevant for globular
cluster dating.

No experiment aimed at a direct determination of  CNO neutrinos from the sun 
has been performed so far. Gallium and chlorine experiments are sensitive 
to neutrinos from the CNO cycle, however these  provide an undistinguishable 
contribution to the total signal, dominated by neutrinos from the $pp$ chain. 
A combined analysis of all solar and reactor neutrino experiments could only provide an upper 
bound for the CNO contribution to solar luminosity, $L(CNO)/L_0 <7.3\%$ at $3\sigma$ \cite{lcno}.

As well known, the key reaction for deciding  the efficiency of the 
CNO cycle  (see Fig. 1) 
is the radiative proton capture  
$^{14}N(p,\gamma)^{15}O$.  It has been measured by several groups in the last fifty years, 
but only the measurements of Schr\"{o}der et al.\cite{schroeder}
 extend over a wide energy range, above $E_{cm}=200$ keV.
For the extrapolation to lower energies,
the effect of  a subthreshold resonance at $E_r=-504$ keV is important.
The zero energy astrophysical $S$-factor  recommended by the NACRE compilation \cite{nacre} 
\begin{equation}
S_{SSM}=(3.2 \pm 0.8) \mbox{ keVb}\quad 
\end{equation}
 is mainly based on data from \cite{schroeder}.
This value, which we shall refer to as the SSM value, agrees  
with the estimate in the compilation by Adelberger et al.\cite{adelberger}, 
~$S=(3.5 ^{+ 0.4} _{ -1.6})$keVb.   

Angulo and Descouvemont \cite{angulo}
reconsidered the data of \cite{schroeder} within the framework of an 
R-matrix model.
The $S$-factor which is extracted from their  analysis is essentially halved with respect to that of
NACRE. In conclusion, whereas all the $S$-factors for the 
$pp$ chain  have been determined with
accuracy of 10\% or better, until recently the key cross section for the CNO cycle 
was  known with an uncertainty of a factor two. 

A new experiment, performed at  the underground Gran Sasso laboratory by the LUNA collaboration, 
has just presented its first results \cite{luna14n}. The new value of the astrophysical $S$-factor
\begin{equation}
\label{S114_luna}
S_{LUNA}=(1.7\pm 0.2) \mbox{keV b} 
\end{equation}
is in good agreement with that found in ref. \cite{angulo}.

The present letter addresses the following questions:

i)what is the impact on solar neutrinos, produced  by  change of a factor  two in 
the astrophysical $S$-factor of  $^{14}N(p,\gamma)^{15}O$?

ii)which are the consequences for the estimated globular cluster ages?

\subsection{The sun and  $^{14}N({\it p},\gamma)^{15}O$ }

As a first approximation, one expects that the produced fluxes 
\footnote{Here and in the following we refer to ``produced fluxes'' and ``produced
signals''  to indicate predictions in the absence of oscillation.}
of CNO neutrinos 
are proportional to 
 the astrophysical $S$-factor, since the production rates  
depend linearly on $S$ 
whereas the density of reacting  nuclei and temperature are weakly dependent on it, 
being essentially fixed by the solar mass, composition and age and  by the main mechanism - the $pp$ chain - 
for energy production.

A decrease by a factor two in $S$ will thus be accompanied by a correponding decrease of the CNO produced fluxes.
Since the total produced neutrino flux is fixed by the observed solar luminosisty, this has to 
be accompanied by an increase of neutrinos from the $pp$ chain.

This qualitative picture is confirmed by the solar models which we have built 
with FRANEC\cite{ciacio}
 for different values of $S$, see Figs. 2 and 3. 
We find a linear dependence of the CNO produced flux which is essentially 
compensated by the variation of  $\Phi(pp)$, the Beryllium flux being practically unchanged. 
The Boron flux  slighlty decreases with decreasing $S$.

The sensitivity of the produced fluxes to  changes of the physical 
and chemical inputs of SSM is usually parametrized in terms of  power laws\cite{beyond03,libro,report}:
\begin{equation}
\label{power}
\Phi(i)=  \Phi_{SSM}(i) (S/S_{SSM} )^{\alpha_i}
\end{equation}

These laws are valid for small changes of the input with respect to the SSM adopted value.
In our case it is preferable to resort to a linear parametrization:
\begin{equation}
\label{linear}
\Phi(i)=  \Phi_{SSM}(i) [ 1 + a_i \frac{(S-S_{SSM})} {S_{SSM}} ]
\end{equation}

The values of $a_i$, as obtained by  
least square fitting, are presented in Table \ref{tabai}.

The produced signal rates  $R_{pr}$ in radiochemical experiments are defined as:
\begin{equation}
\label{prod_signal}
 R_{pr} = \Sigma_i \sigma(i) \Phi(i)
\end{equation}
where $\sigma(i)$ is the  capture cross section of the $i$-th neutrinos in the detector.
Their dependence on $S$ is shown in Tables \ref{tabmodelli1} and \ref{tabmodelli2} and Fig. 4.
One sees that the linear $S$-dependence of the fluxes translates into a similar behaviour of the signals.
For gallium detectors, the reduction of $S$ correponds to the fact that CNO neutrinos 
are replaced with a similar  number of lower energy pp neutrinos. 
Since the cross section decreases when energy decreases, the resulting signal is smaller.
We remind that pp neutrinos are below threshold for chlorine detectors, 
sensitive to B, Be and CNO neutrinos. For this reason, a decrease of CNO neutrinos 
results in a signal decrease.

These effects are enhanced  when considering the effective signal rates, 
which include neutrino oscillations,
\begin{equation}
\label{eff_signal}
 R = \Sigma_i \sigma(i) \Phi(i) P_{ee}(i)
\end{equation}
where $P_{ee}(i)$ is the average survival probability of $i$-th neutrinos.
The LMA solution predicts   $P_{ee}$ decreasing  with energy, 
so that the  percentage contribution of lower 
energy neutrinos is larger in the effective signal than  in the 
produced signal.

For gallium and chlorine experiments the change $S_{SSM} \rightarrow S_{LUNA}$
results in a signal reduction $\Delta R(Ga)=2$ SNU and $\Delta R(Cl)=0.1$ SNU
This alleviates the slight tension between the chlorine result
and the LMA+SSM prediction noticed in \cite{smirnov} and \cite{roadmap}.

Complementary to neutrinos, helioseismic observations provide us with a detailed view of the solar interior.
By means of helioseismology one can reconstruct   the sound speed profile inside the sun. 
Recent SSMs, all including gravitational settling and elemental diffusion, 
are well in agreement with helioseismic data, see e.g. \cite{eliosnoi}.
Since the efficiency of the CNO cycle is important for energy generation in  
the innermost part of the sun, in this region  one expects that the sound speed  
is sensitive to $S$.

In Fig. 5  we compare solar models  computed with different  $S$ values.
We find that all models with  $S<S_{SSM}$  cannot be distinguished from the SSM 
within the uncertainties of the helioseismic observables.
On the other hand, models with $S>5 \, \cdot  S_{SSM}$ are excluded at $3\sigma$ ore more.
In other words, the helioseismic limit to the energy generation rate by the CNO cycle in the sun is
$L(CNO) / L_0 <7.5 \%$, comparable to the bound obtained in \cite{roadmap} from solar neutrino experiments.

\subsection{Globular clusters and $^{14}N({\it p},\gamma)^{15}O$ }

The age of globular clusters, the oldest objects  in the Galaxy,
is extremely important for understanding the
galactic evolution and to give a firm lower limit to the 
Galaxy formation epoch.


The effect of the  $^{14}N(p,\gamma)^{15}O$ cross section on the 
evolutionary characteristics of population II stars
has been analysed  in \cite{Brocato98}.  
In order to assess the impact of the LUNA result, 
we repeat and extend the calculations of \cite{Brocato98} by means of a FRANEC version
containing  updated physical inputs, see \cite{Cassisi99,Cariulo03}, and
including microscopic diffusion of He and heavy elements. 
The external convection is treated within  a
mixing length approach and its efficiency  is calibrated so as to reproduce the red
giant branch (RGB) color of globular clusters with  different chemical
composition. The resulting models 
reproduce the color-magnitude diagram of well
observed globular clusters as M68, M3 and M13 \cite{Cariulo03}.

The present calculations have been made for two chemical compositions
(Z=0.0002, Y=0.230 and Z=0.001, Y=0.232) which are well representative
of the globular cluster population.

As well known, the main parameter marking the age of a stellar cluster
is the luminosity at  Turn-Off, $L_{\rm TO}$,  in the Hertzprung-Russell (HR) diagram,
 see e.g. \cite{Renzini91}.
Stellar evolution theory can predict the behaviour 
of $ L_{\rm TO}$  as a function of cluster  age $t$.
Clearly, the relationship between $L_{\rm TO}$ and $t$ depends on the physical inputs 
adopted in the calculations and in particular on the assumed value of $S$.

In Fig. 6  we show  the  12 Gyr
isochrones  obtained for different 
values of the $^{14}N(p,\gamma)^{15}O$ astrophysical  factor, with  the same chemical
composition (Z=0.001, Y=0.232).
By decreasing $S$, $ L_{\rm TO}$  increases. 
In fact the Turn-Off (TO) marks the onset of the CNO hydrogen 
burning in shells.
If the CNO efficiency is reduced, this onset is delayed and
 TO occurs at a later time and with a larger luminosity \cite{Brocato98}.

The dependence  is shown more quantitatively in Fig.7 for the higher metallicity
composition (the low metallicity case looks similar).
By halving  $S$ the same value of $ L_{\rm TO}$
corresponds to an age increase\footnote{A similar 
conclusion has been obtained  by O. Straniero et al. in \cite{oscar}.}
$\Delta t \simeq 0.7$ Gyr.

This approach assumes that $L_{\rm TO}$ can be fixed from observations,
independently of $S$. Actually the determination of  $L_{\rm TO}$
requires the knowledge of the  cluster distance modulus, which is
often  obtained by using  
RR Lyrae stars in the Horizontal Branch (HB)
\footnote{For the sake of precision, our candles are provided 
by the HB lower envelope (Zero Age Horizontal Branch, ZAHB)
in the  RR Lyrae region.}
as standard candles.
In this case the relevant observable is the ratio of 
$ L_{\rm TO}$ to the HB luminosity, $L_{\rm HB}$,
which is independent of the cluster distance.
A frequently used variable for determining the cluster age
 is defined as:
\begin{equation}
\log L_{\rm HB-TO}=
\log (L_{\rm HB} / L_{\rm TO}) \quad . 
\end{equation}

As discussed in \cite{Brocato98}, variations of $S$
also affect the HB stars, which are powered by He burning in
the core and by H burning 
in a  surrounding shell, mainly through the CNO cycle. 

A decrease of $S$ has two competing  effects:
 it decreases the  CNO cycle efficiency, driving
a decrease of $L_{\rm HB}$, and at the same time it
produces an increase of the helium core mass 
at  He ignition, which reflects into an increase 
of $L_{\rm HB}$. 
The net effect depends on the cluster metallicity. 
For low metallicities ($Z=0.0002$) 
we find that a decrease of $S$ by a factor two 
leads to an increase 
$\Delta \log L_{\rm HB}\sim0.01$.
For moderately metal-rich HB stars ($Z=0.001$), 
where  CNO burning is more important, the same variation produces 
instead a decrease $\Delta \log L_{\rm HB}$ of about the same amount.
This means that when using  $\log L_{\rm HB-TO}$ as age indicator
the  LUNA result leads to an increase of the
estimated age which depends on the cluster metallicity:
we obtain $\Delta t \sim 0.5$ Gyr for $Z=0.0002$ and 
$\Delta t \sim 1$ Gyr 
for $Z=0.001$.

The determination of globular clusters ages is
presently affected by several uncertainties, resulting
from  the chemical composition, 
from the adopted  physical inputs and 
from the efficiency of various physical mechanisms (e.g. microscopic diffusion). 
Additional uncertainties arise from the comparison between
theoretical and observed luminosities,
see e.g. \cite{Chaboyer95,Chaboyer96,Chaboyer98,Castellani99}.

A precise determination of the overall uncertainty is thus difficult.
The cluster age as determined from the absolute value of  
$ L_{\rm TO}$  (i.e. assuming that the cluster distance  is known in an 
indipendent way) is affected by an error of  
$\sim 1.5$ Gyr \cite{Cassisi99,Castellani99,Chaboyer95}.
If  $\log L_{\rm HB-TO}$ is used as an age indicator, the uncertainty  is  $\sim 2.0$ Gyr,
see \cite{Cassisi99,Castellani99}. 
The increase of the globular cluster ages following from LUNA result is thus 
within the error bar of
the present determinations.  Nevertheless the new and more precise value of $S$ will
 be important when better astrophysical inputs will be available. 

We conclude this section by discussing the  effect 
 on another interesting evolutionary feature of globular clusters: 
the so called RGB bump,  a region of the HR cluster 
diagram with  higher star density.
The RGB bump corresponds the momentanous decrease of
the stellar luminosity in RGB which marks the encounter of the 
CNO H-burning shell with the chemical discontinuity produced by 
the first dredge-up, see e.g. \cite{Thomas67,Iben68,Renzini88}. 
We find that a reduction of  $S$
by a factor two leads to an increase of the bump
luminosity of about $\Delta \log L_{\rm bump} \sim 0.02$.
which is well within the estimated theoretical and observational 
uncertaintes on this quantity, see e.g. 
\cite{Cassisi97,Cass_Degli97,Ferraro99}.

\subsection{Concluding remarks}
We summarize here the main points of this paper:
\begin{itemize}

\item
The LUNA result on the astrophysical $S$-factor for  $^{14}N(p, \gamma)^{15}O$ 
implies that SSM+LMA predictions for the gallium and
chlorine experiments are reduced by 2 and 0.1 SNU respectively.
This alleviates the slight tension between theory and chlorine result.

\item
The new $S$ value does not change significantly helioseismic
observables.

\item
On the other hand, helioseismology excludes a  CNO
contribution to solar luminosity larger than 7.5\%.

\item
The age of globular clusters is increased by a quantity
0.5--1 Gyr, depending on the method for determining the Turn-Off
luminosity and on the cluster metallicity.

\end{itemize}

~\\
We are extremely grateful to C. Broggini,
V. Castellani, H.P. Trautvetter and C. Rolfs for useful discussions.

~\\
This work was performed within the Astroparticle Physics project
financed by MIUR as PRIN-2002.


\begin{table}
\caption{\bf Slope of the flux dependence on $S$.}
\begin{tabular}{cc}
Source	& 	$a_i$	\\
\hline
pp & -0.013  \\
pep & -0.018  \\
Be & -0.003  \\
B & +0.018  \\
N & +0.875  \\
O & +1.008  \\
\end{tabular}
\label{tabai}
\end{table}

\begin{table}
\caption{{\bf Produced fluxes and signals with $S_{SSM}=3.2$ keVb}. 
All other inputs as in [18].}
\begin{tabular}{llll}
	&	Flux 	&  Cl	         & Ga	\\

source & [$10^{9}$cm$^{-2}$s$^{-1}$]  & [SNU] &  [SNU] \\
\hline
pp 	& 59.99 	& 0 		& 70.3 \\
pep 	& 0.142  	& 0.227         & 2.89 \\
Be 	& 4.52	 	& 1.08		& 32.4 \\	
B	& 5.21 $10^{-3}$ & 5.94 	& 12.5 \\
N	& 0.515		& 0.0875	& 3.13 \\
O	& 0.437		& 0.297 	& 4.97 \\
\hline 
total	&		& 7.64 		& 126.3 \\
\end{tabular}
\label{tabmodelli1}
\end{table}

\begin{table}
\caption{\bf{Produced fluxes and signals when $S_{SSM} \rightarrow S_{LUNA}$}}
\begin{tabular}{llll}
	&	Flux 	&  Cl	& Ga	\\

source & [$10^{9}$cm$^{-2}$s$^{-1}$]  & [SNU] &  [SNU] \\
\hline
pp 	& 60.33         & 0 		& 70.7 \\
pep 	& 0.143		& 0.229         &  2.92 \\
Be 	& 4.53	 	& 1.09		& 32.5 \\	
B	& 5.17 $10^{-3}$ & 5.90 	& 12.4 \\
N	& 0.305		& 0.0518	& 1.84 \\
O	& 0.226		& 0.154		& 2.57 \\
\hline 
total	&		&  7.43		&  123.2 \\
\end{tabular}
\label{tabmodelli2}
\end{table}


\begin{figure}[tbh]
\begin{center}
\label{figcno}
  \epsfig{figure=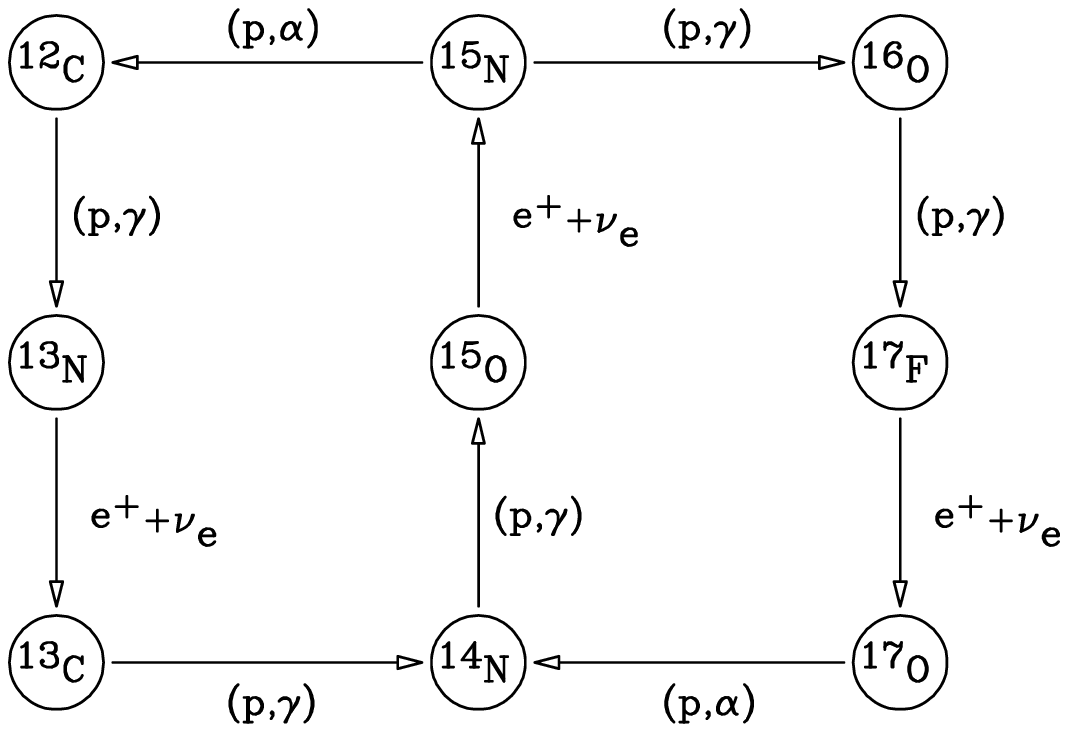,width=0.9\hsize,angle=90}
\vspace{2cm}
\end{center}
\caption[b]{ The CNO bi-cycle. }
\end{figure}

\begin{figure}[tbh]
\begin{center}
  \epsfig{figure=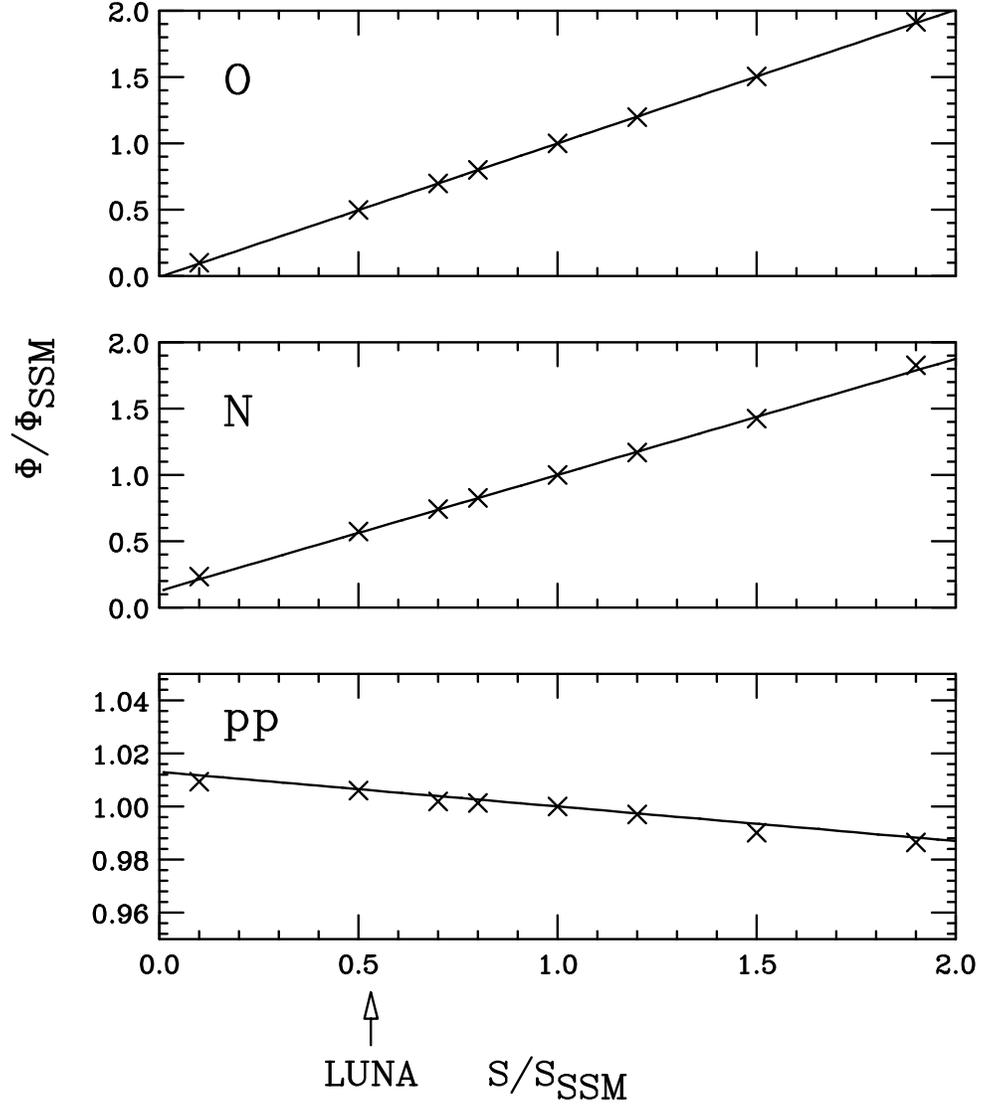,width=0.9\hsize,angle=90}
\vspace{2cm}
\caption[b]{{\bf  Produced O, N and pp fluxes.}
Crosses denote solar model results and the straight lines are  linear fits.
We use $S_{SSM}=3.2$ keVb. The arrow
corresponds to the new value $S_{LUNA}=1.7$ keVb.}
\end{center}
\label{figflussi1}
\end{figure}

\begin{figure}[tbh]
\begin{center}
  \epsfig{figure=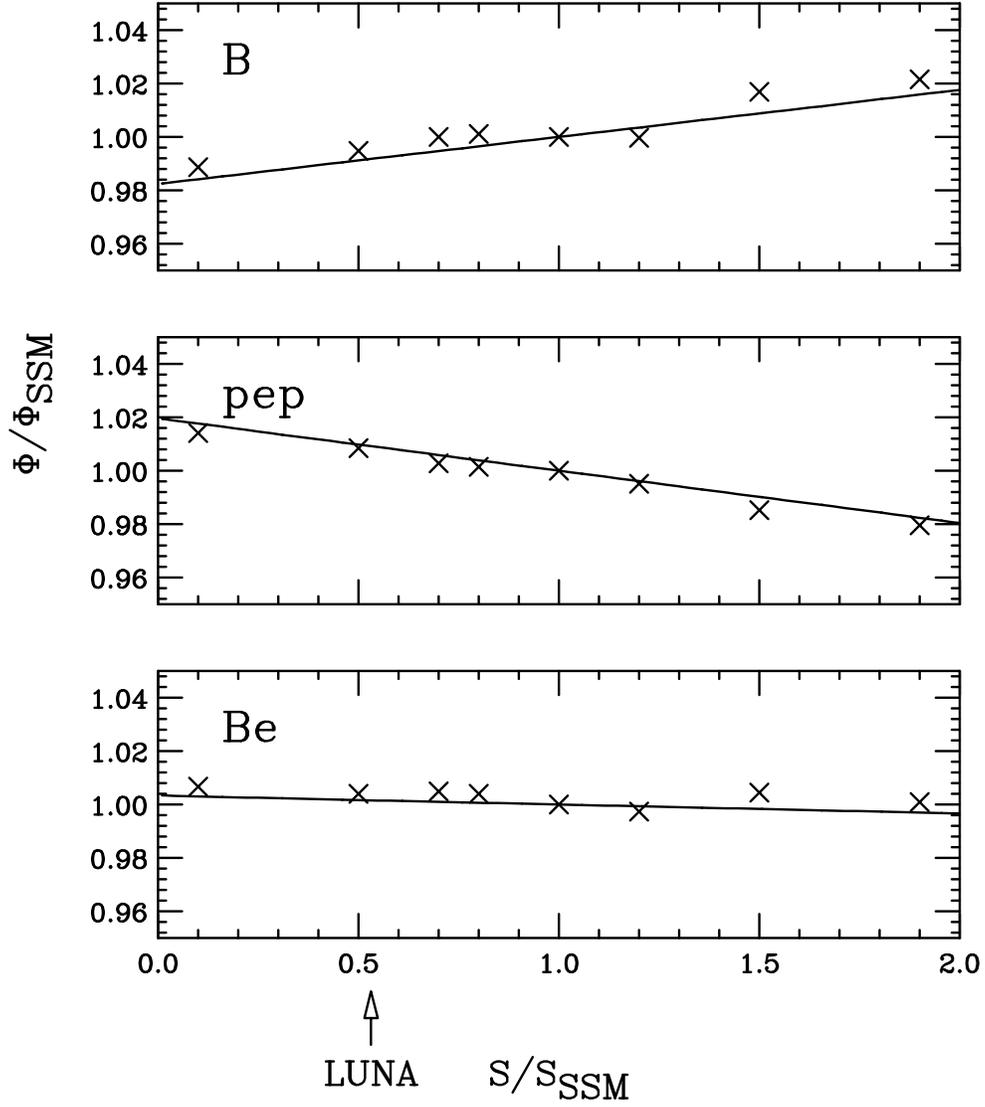,width=0.9\hsize,angle=90}
\vspace{2cm}
\caption[b]{{\bf Produced B, pep and Be fluxes.} Same notation as in Fig. 2.}
\end{center}
\label{figflussi2}
\end{figure}

\begin{figure}[tbh]
\begin{center}
  \epsfig{figure=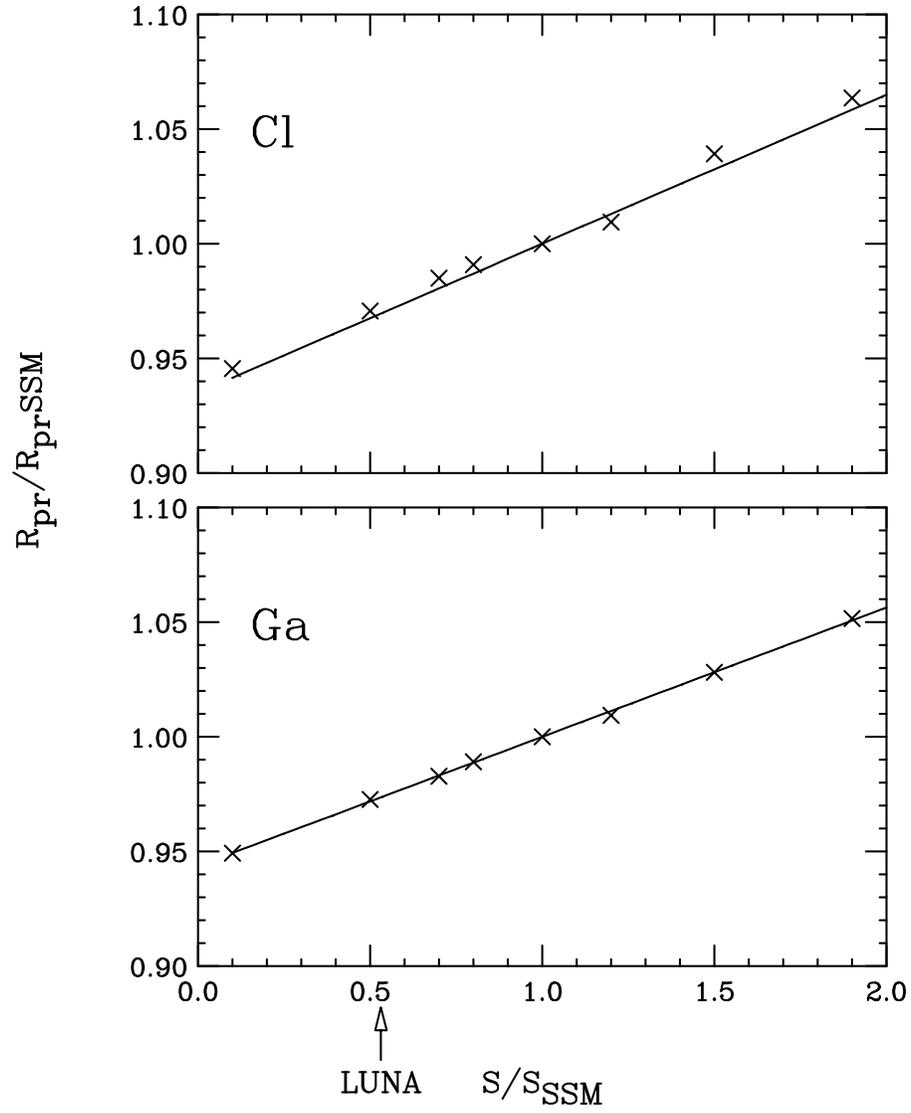,width=0.9\hsize,angle=90}
\vspace{2cm}
\caption[b]{{\bf Produced signals in radiochemical experiments}.  Same notation as in Fig. 2.}
\end{center}
\label{figsegnali}
\end{figure}

\begin{figure}[tbh]
\begin{center}
  \epsfig{figure=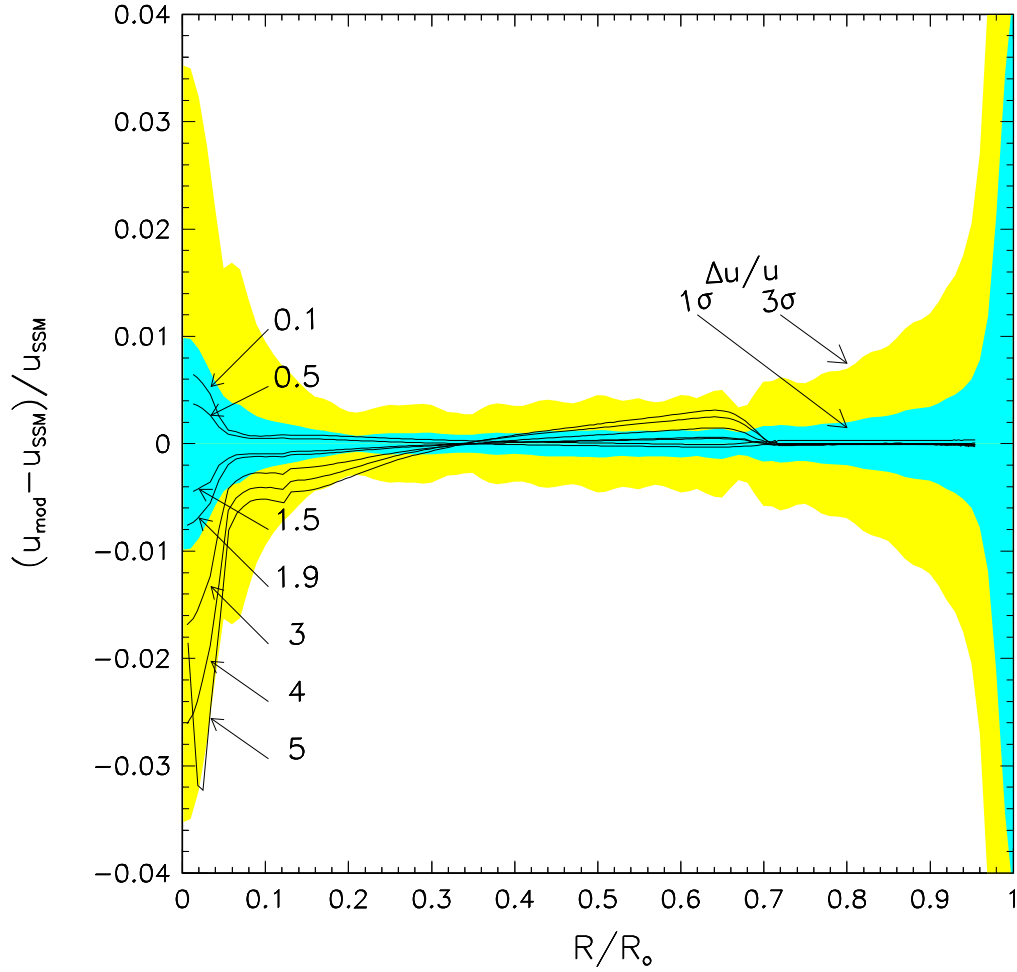,width=0.9\hsize}
\caption[b]{
Relative change (model-SSM)/SSM  of the squared isothermal sound speed $u=P/\rho$ 
as a function of the radial coordinate, for the indicated values of $S/S_{SSM}$.
The dark (light) shaded area corresponds to the 
$1\sigma \, (3\sigma)$ uncertainty on helioseismic determination \cite{eliosnoi}. 
}
\end{center}
\label{figelio}
\end{figure}

\begin{figure}[tbh]
\begin{center}
  \epsfig{figure=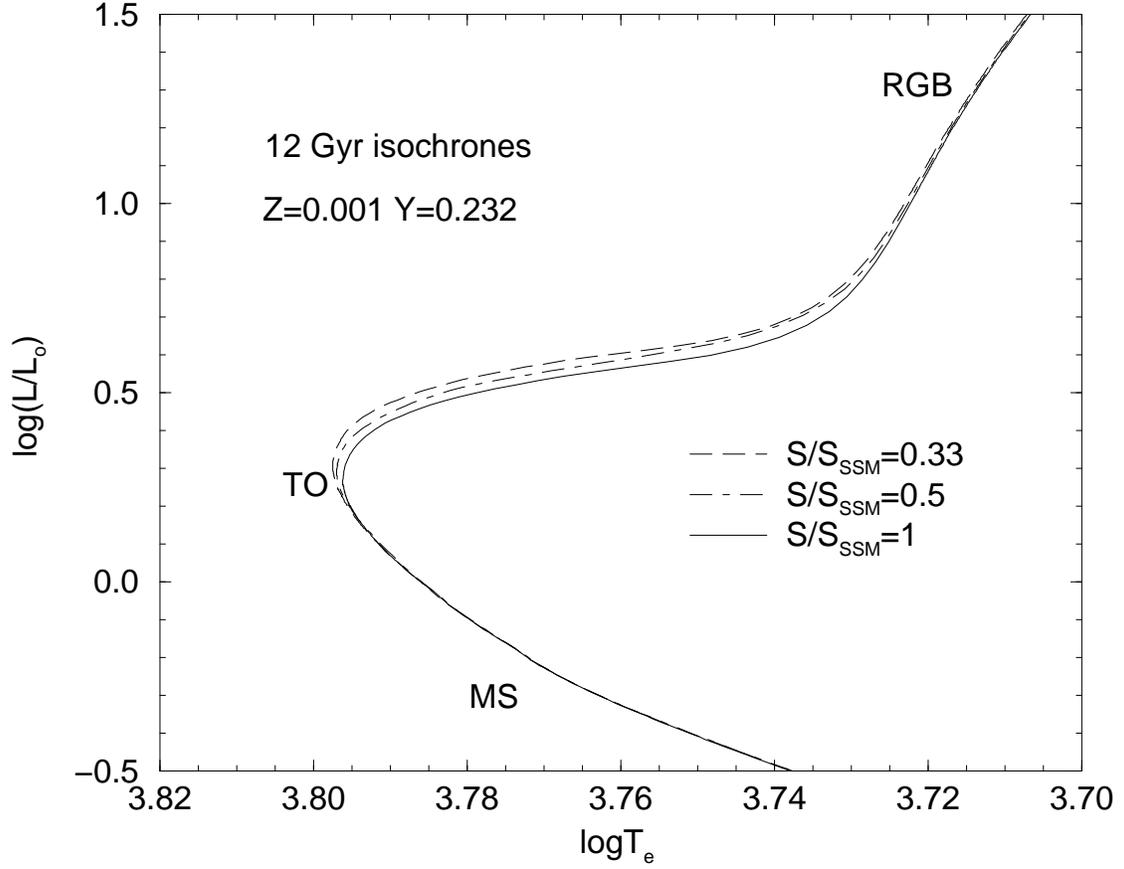,width=0.9\hsize}
\caption[b]{
{\bf Isochrone dependence on $S$}. The calculated luminosity $L$, in units
of the solar luminosity $L_0$, is presented as a function of
the effective temperature T$_e$ in Kelvin.
}
\end{center}
\label{figGCA}
\end{figure}

\begin{figure}[tbh]
\begin{center}
  \epsfig{figure=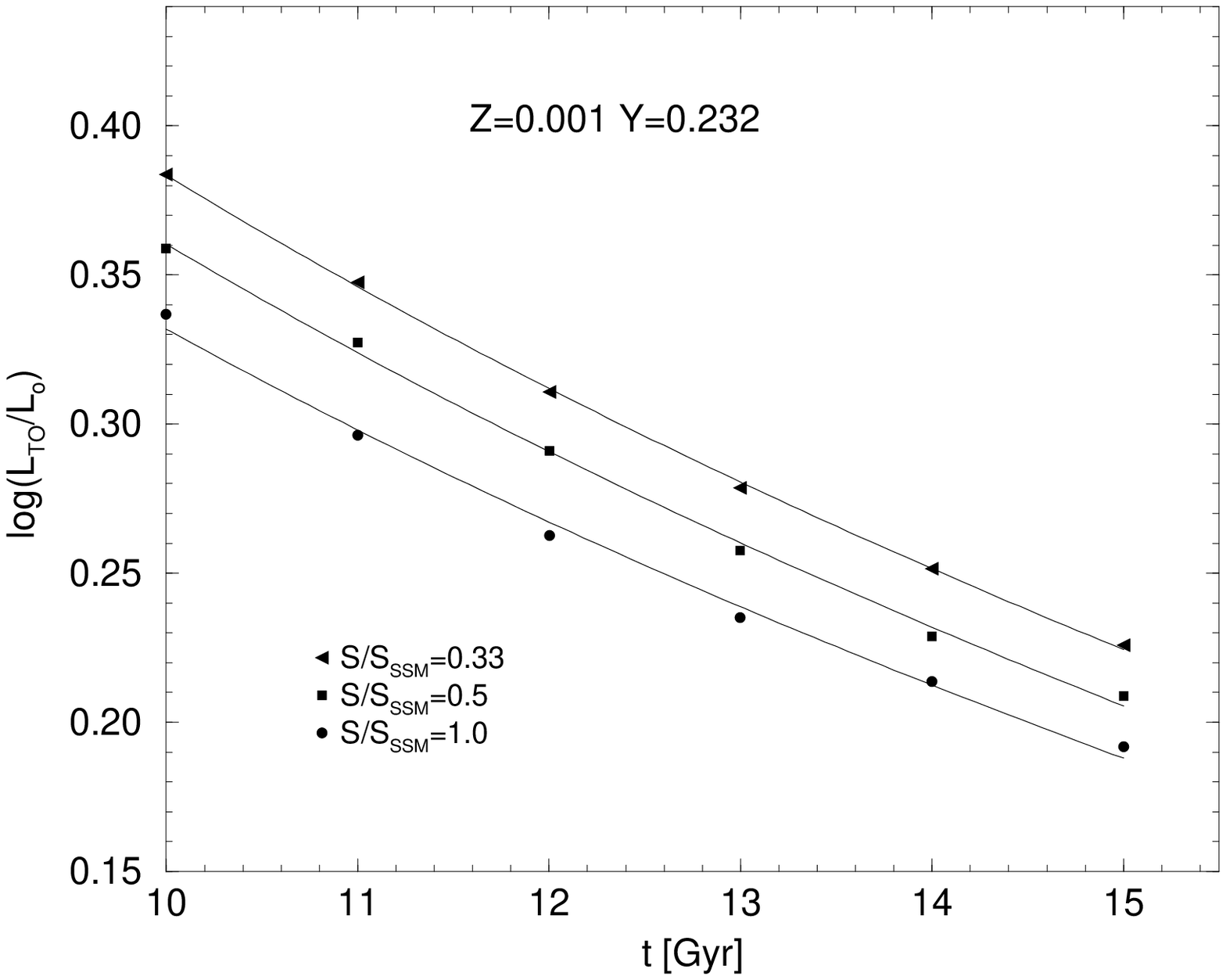,width=0.9\hsize}
\vspace{3cm}
\caption[b]{
{\bf Luminosity at Turn-Off $L_{\rm TO}$ as a function of the
cluster age $t$}. Points are the results of evolutionary calculations, 
continuous curves correspond to linear fits in the $\log L_{\rm TO} -\log t$ plane.
}
\end{center}
\label{figGCB}
\end{figure}


\begin{thebibliography}{99}

\bibitem{sno}SNO coll., nucl-ex/0309004.
\bibitem{kamland}K. Eguchi et al., Phys. Rev. Lett. 90 (2003) 021802.
\bibitem{roadmap} J.N. Bahcall and C. Pena-Garay, JHEP 0311 (2003) 004

\bibitem{smirnov}P.C. de Holanda and  A.Yu. Smirnov, hep-ph/0309726.


\bibitem{eddington}S.A. Eddington, Nature 106 (1920) 14.

\bibitem{bethe} H.A. Bethe, Phys. Rev. 55 (1939) 103 and 434.
\bibitem{von} C.F. von Weizs\"{a}cker, Phy. Z. 39 (1938) 633.
\bibitem{beyond03}G. Fiorentini and B. Ricci, astro-ph/0310753.
\bibitem{luna33} M. Junker et al. (LUNA coll.), Phys. Rev. C 57 (1998) 2700.
\bibitem{lcno} J.N. Bahcall, M.C. Gonzales-Garcia and C. Pe\~{n}a-Garay, Phys. Rev. Lett. 90 (2003) 131301


\bibitem{schroeder} U. Schr\"{o}der et al, Nucl. Phys. A 467 (1987) 240.
\bibitem{nacre} C. Angulo et al.(NACRE coll.), Nucl. Phys. A 656 (1999) 3.
\bibitem{adelberger} E.G. Adelberger et al., Rev. Mod. Phys. 70 (1998) 1265.
\bibitem{angulo} C. Angulo and P. Descouvemont Nucl. Phys. A 690 (2001) 755.
\bibitem{luna14n}A. Formicola et al. (LUNA coll.), nucl-ex/0312015
\bibitem{ciacio}F. Ciacio, S. Degl'Innocenti and B. Ricci, Astr. Astroph. Suppl. Ser. 123 (1997)449.
\bibitem{libro}J.N. Bahcall, ``Neutrino Astrophysics'', Cambridge University Press, Cambridge 1989.
\bibitem{report}V. Castellani et al., Phys. Rep. 281 (1997) 310.
\bibitem{eliosnoi} W. J. Dziembowski et al., Astr. Phys. 7 (1997) 77

\bibitem{Brocato98}E. Brocato, V. Castellani and F. Villante, MNRAS 298 (1998)  557
\bibitem{oscar}O. Straniero et al., in 
Proc. of the 11th Workshop on Nuclear      
Astrophysics, Ringberg Castle (Germany) February 11 - 16, 2002,
 MPA/P13, Garching b. Munchen: Max Planck Institut fur Astrophysik, 2002,  
p.30-36, W. Hillebrandt and E. M\"uller (Eds.)                                   


\bibitem{Cassisi99}S. Cassisi et al. A\&AS 134 (1999) 103   

\bibitem{Cariulo03}P. Cariulo, S.Degl'Innocenti and V.Castellani, A\&A (2003) submitted
\bibitem{Renzini91}A. Renzini, 1991, in ``Observational tests of 
cosmological inflation'',T. Shanks, A. J. Banday and  R. S. Ellis (Dordrecht: Kluwer), 131
\bibitem{Chaboyer95}B. Chaboyer ApJ 444 (1995) L9
\bibitem{Chaboyer96}B. Chaboyer et al. MNRAS 283 (1996) 683
\bibitem{Chaboyer98}B. Chaboyer et al. ApJ 494 (1998) 96
\bibitem{Castellani99}V. Castellani and S. Degl'Innocenti A\&A 344 (1999) 97


\bibitem{Cassisi97}S. Cassisi and M. Salaris MNRAS 285 (1997) 593
\bibitem{Cass_Degli97}S. Cassisi, S. Degl'Innocenti and M. Salaris MNRAS 290 (1997) 515
\bibitem{Ferraro99}F. Ferraro et al. AJ 118 (1999) 1738
\bibitem{Thomas67}H.C. Thomas, Z. Astrophys. 67 (1967) 420
\bibitem{Iben68}I. Iben Jr. Nature 220 (1968) 143
\bibitem{Renzini88}A. Renzini and F. Fusi Pecci,  ARA\&A 26 (1988) 199




\end{thebibliography}
\end{document}